\documentclass{emulateapj}

\slugcomment{Accepted for publication in The Astrophysical Journal Letters}
\shorttitle{Tidal Interactions in Merging White Dwarf Binaries} 
\shortauthors{PIRO}

\newcommand{\be}{\begin{eqnarray}}
\newcommand{\ee}{\end{eqnarray}}
\newcommand{\lp}{\left(}
\newcommand{\rp}{\right)}

\begin{document}


\title{Tidal Interactions in Merging White Dwarf Binaries}

\author{Anthony L. Piro}

\affil{Theoretical Astrophysics, California Institute of Technology, 1200 E California Blvd., M/C 350-17, Pasadena, CA 91125; piro@caltech.edu}


\begin{abstract}
The recently discovered system J0651 is the tightest known detached white dwarf (WD) binary. Since it has not yet initiated Roche-lobe overflow, it provides a relatively clean environment for testing our understanding of tidal interactions. I investigate the tidal heating of each WD, parameterized in terms of its tidal $Q$ parameter. Assuming that the heating can be radiated efficiently, the current luminosities are consistent with $Q_1\approx7\times10^{10}$ and $Q_2\approx2\times10^7$, for the He and C/O WDs, respectively. Conversely, if the observed luminosities are merely from the cooling of the WDs, these estimated values of $Q$ represent upper limits. A large $Q_1$ for the He WD means its spin velocity will be slower than that expected if it was tidally locked, which, since the binary is eclipsing, may be measurable via the Rossiter-McLaughlin effect. After one year, gravitational wave emission shifts the time of eclipses by $5.5\ {\rm s}$, but tidal interactions cause the orbit to shrink more rapidly, changing the time by up to an additional $0.3\ {\rm s}$ after a year. Future eclipse timing measurements may therefore infer the degree of tidal locking.
\end{abstract}



\section{Introduction}
\label{sec:intro}

Compact white dwarf (WD) binaries are important as progenitors for a variety of interesting astrophysical systems and/or events. Although the merger is poorly understood, it may result in a helium-rich sdB star \citep{sj00,sj02,heb09}, R CrB star \citep{web84}, survive as mass-transferring AM CVn binaries \citep{py06} which may produce a .Ia supernova \citep{bil07}, or even result in a class of Type Ia supernovae \citep{it84}. Prior to Roche-lobe overflow, these are among the strongest gravitational wave sources in our Galaxy \citep{nel09}. Tidal interactions are important at these short distances, which may alter the inspiral and determine the fate upon merger. Unfortunately, most tight binaries are also accreting \citep[as studied in][]{mar04,del05}, which complicates our ability to isolate the effects of tides.

The detached WD binary SDSS J065133.33+284423.3 \citep[hereafter J0651;][]{bro11} has an orbital period $P=765\ {\rm sec}$, the smallest of any such binary yet discovered. Furthermore, the system's lightcurve shows ellipsoidal variations, Doppler boosting, and primary and secondary eclipses, which allow the properties of this system to be tightly constrained. Since there is no accretion, J0651 is ideal for studying the role of tides.

In the following study I explore the effects of tidal interactions in J0651, focusing on tidal heating and eclipse measurements. In \S \ref{sec:equations} I summarize the tidal and orbital evolution equations and estimate the asynchronicity and tidal heating expected. In \S \ref{sec:calculate} I follow the orbital evolution with time. I summarize how tides cause deviations in the period derivative and eclipse timing from what is expected if the evolution was driven merely by gravitational waves. This is in stark contrast to the Hulse-Taylor pulsar, which has a period derivative exactly equal to that predicted from Einstein's theory \citep{ht75,wei10}. In \S \ref{sec:conclusion} I summarize the results and discuss  future work that can be done on this problem.

\section{Governing Equations and Analytic Estimates}
\label{sec:equations}

I begin by summarizing the equations used to study the time evolution of the binary. Following the convention in \citet{bro11}, I refer to the less massive He WD as the primary, with $M_1=0.25M_\odot$ and $R_1=0.035\ R_\odot$, and the more massive C/O WD as the secondary, with $M_2=0.55\ M_\odot$ and $R_2=0.013\ R_\odot$.

The tidal force raises a tide on the primary, $H_1$, inducing a quadrupole moment
\be
 	U_{\rm grav, 1}\sim \lp \frac{H_1}{R_1}\rp M_1R_1^2\sim M_2R_1^2\lp \frac{R_1}{a}\rp^3,
\ee
where $a$ is the orbital separation\footnote{For these derivations I focus on the primary, but these equations can just as well apply to the secondary by switching the subscripts 1 and 2.}. The component of the quadrupole that contributes to secular changes in the orbit is \citep{gs66},
\be
	U_{\rm grav, 1}^{\rm sec} \sim Q_{\rm grav}\frac{\sigma_1}{n}\sin\epsilon_1
	\sim\frac{M_2 R_1^2}{Q_1} \frac{\sigma_1}{n} \lp\frac{R_1}{a}\rp^3
\ee
where $\epsilon_1\sim1/Q_1$ is the angle of lag due to internal friction, $\sigma_1= 2(n-\Omega_1)$ is the tidal forcing frequency in the rotating frame of the primary (the factor of 2 is due to the $m=2$ perturbation represented by the tidal deformation), and $n=2\pi/P$ is the orbital frequency.

The tidal $Q$ that is appropriate for WD binaries is not currently known. In other astrophysical systems, $Q$ can vary quite dramatically from $\sim 10$ for rocky bodies like the Earth \citep{gs66} to $\sim10^5-10^6$ for gaseous planets such as Jupiter or extrasolar Jupiters \citep{pg80,ol04,wu05}. For WD binaries, \citet{cam84} estimated synchronization times of $\sim10^8\ {\rm yr}$, which correspond to $Q\sim10^{12}$. \citet{wil10} found $Q\sim10^{15}$, but their analysis did not include dynamical tide effects, so this is merely an upper limit. \citet{fl11} looked at dynamical tides and then, appealing to \citet{wu98} for damping times, estimate $Q\sim10^4-10^{11}$ (depending the radial wavelength of the dominant excited mode). Other studies, such as \citet{rac07}, are more focused on when accretion is occurring. Due to these uncertainties, I take $Q$ to be a free parameter for the present study.

This secular quadrupole moment implies a torque
\be
	N_1 \sim U_{\rm grav, 1}^{\rm sec} n^2 \sim \frac{M_2 R_1^2}{Q_1} \lp\frac{R_1}{a}\rp^3 n\sigma_1.
	\label{eq:torque}
\ee
Assuming solid-body rotation, the spin evolves as
\be
	\frac{d}{dt}\lp I_1\Omega_1 \rp \sim N_1 \sim \frac{M_2 R_1^2}{Q_1} \lp\frac{R_1}{a}\rp^3 n\sigma_1,
	\label{eq:angularmomentum}
\ee
where $I_1=kM_1R_1^2$ and I set $k=0.2$ \citep[][but this may be modified for the hot He WD]{mar04}.

The angular momentum of the binary evolves as
\be
	\dot{J}_{\rm orb} = \dot{J}_{\rm gw} - N_{1} - N_2
	\label{eq:gw}
\ee
where the angular momentum loss to gravitational waves is \citep{ll75}
\be
	\dot{J}_{\rm gw} = -\frac{32}{5}\frac{G^3}{c^5}\frac{M_1M_2M}{a^4}J_{\rm orb},
\ee
where $J_{\rm orb} = (Ga/M)^{1/2}M_1M_2$ and $M=M_1+M_2$.

The evolution depends on two key timescales. The first is the gravitational wave timescale,
\be
	\tau_{\rm gw} &=& P\left|\frac{dP}{dt}\right|^{-1}= \frac{5}{96} \frac{c^5}{G^{5/3}} \frac{M^{1/3}}{M_1M_2}\lp\frac{P}{2\pi}\rp^{8/3}
	\nonumber
	\\
	&=& 5.8\times10^6 M_{0.25}^{-1}M_{0.55}^{-1}M^{1/3}_{0.8} P_3^{8/3}\ {\rm yr},
	\label{eq:taugw}
\ee
where $M_{0.25}=M_1/0.25M_\odot$, $M_{0.55}=M_2/0.55M_\odot$, $M_{0.8}=M/0.8M_\odot$ and $P_3 = P/10^3\ {\rm s}$. The other is the torquing timescale found from equation~(\ref{eq:angularmomentum}),
\be
	\tau_1 &=& \sigma_1 \lp\frac{d\Omega_1}{dt}\rp^{-1}=  kQ_1\lp\frac{M_1}{M_2} \rp \lp \frac{GM}{R_1^3}\rp \lp\frac{P}{2\pi} \rp^3,
	\nonumber
	\\
	&=& 1.6\times10^6k_{0.2} Q_{10}M_{0.25}^{-1}M_{0.55}M_{0.8}R_{9.3}^{-3}P_3^3\ {\rm yr},
\ee
where $R_{9.3}=R_1/2\times10^9\ {\rm cm}$, $k_{0.2} = k/0.2$, and $Q_{10} = Q_1/10^{10}$. In previous calculations, such as \citet{mar04}, the strength of the tidal torquing is discussed in terms of a tidal synchronization timescale, $\tau_s$. This is merely a factor of 2 smaller than the $\tau_1$ defined here. This demonstrates that if $Q_1$ is assumed constant, the synchronization timescale decreases rapidly as the binary shrinks.

Given these timescales, I estimate the asynchronicity of the primary's spin. The time derivative of the tidal forcing frequency is
\be
	\frac{d\sigma_1}{dt} = 2\lp \frac{dn}{dt} - \frac{d\Omega_1}{dt} \rp = 2 \lp \frac{n}{\tau_{\rm gw}} -  \frac{\sigma_1}{\tau_1} \rp.
\ee
The competing effects of the gravitational wave emission promoting asynchronicity and the torques tidally locking the stars drive the system toward an equilibrium where $d\sigma_1/dt~\approx~0$. The tidal forcing frequency is then
\be
	\frac{\sigma_1}{n} & \approx &\frac{\tau_1}{\tau_{\rm gw}} = \frac{96}{5}\frac{G^{8/3}}{c^5}\frac{M_1^2M^{2/3}}{R_1^3}\lp\frac{P}{2\pi}\rp^{1/3}kQ_1
	\nonumber
	\\
	&=&0.27 k_{0.2} Q_{10}M_{0.25}^2M_{0.8}^{2/3}R_{9.3}^{-3} P_3^{1/3}.
	\label{eq:eq}
\ee
The general trend is that as the He WD gets closer to the C/O WD, the tidal torques become stronger (since $\tau_1\propto P^3$ and $\tau_{\rm gw}\propto P^{8/3}$), which in turn makes $\sigma_1/n$ smaller. On the other hand, for a larger $Q_1$ there is less of a lever arm for torquing the He WD, and its spin becomes more asynchronous.

A tidal torque acting on an asynchronously spinning WD implies that work is being done. The rate of energy input is
\be
	\dot{E}_1 &\approx& \sigma_1 N_1 \approx \frac{M_2 R_1^2}{Q_1}\lp \frac{R_1}{a} \rp^3n \sigma_1^2.
	\nonumber
	\\
	 &=& 2.3\times10^{31} k_{0.2}^2 Q_{10}M_{0.25}^4M_{0.55}
	 \nonumber
	 \\
	 &&\times M_{0.8}^{1/3}R_{9.3}^{-1}P_3^{-13/3}{\rm ergs\ s^{-1}}.
	 \label{eq:edot}
\ee
Whether or not this $\dot{E}_1$ is observed depends on how shallow it is deposited. One expects that the luminosity of the He WD is roughly given by $\dot{E}_1$ when the thermal timescale down to where the tide is damped is less than $\tau_{\rm gw}$. \citet{it98} consider this issue in more detail by distributing the tidal heating evenly throughout the WD, and then following the WD's cooling. \citet{bro11} estimate the effective temperatures for the primary and secondary WDs to be $16,400\ {\rm K}$ and $9,000\ {\rm K}$, respectively, providing luminosities of $L_1 = 3.1\times10^{32}\ {\rm ergs\ s^{-1}}$ and $L_2 = 3.8\times10^{30}\ {\rm ergs\ s^{-1}}$. Using equation (\ref{eq:edot}), $Q_1\approx5\times10^{10}$ is needed for $\dot{E}_1\approx L_1$. Equation (\ref{eq:edot}) appears to imply that the heating increases indefinitely with larger $Q_1$, but the equilibrium condition only applies for $\tau_1\lesssim \tau_{\rm gw}$. Using a maximum asynchronicity frequency of $\sigma_1=2n$, I find
\be
	\dot{E}_{\rm 1,max} \approx  5.9\times10^{32} Q_{10}^{-1}M_{0.25} M_{0.8}^{-1}R_{9.3}^{5}P_3^{-5}\ {\rm ergs\ s^{-1}},
	\nonumber
	\\
	 \label{eq:edotmax}
\ee
for the maximum heating rate.

The above estimates focus on the asynchronicity of the He WD, but the same arguments can be extended to the C/O WD. Using equation (\ref{eq:eq}),
\be
	\frac{\sigma_2}{\sigma_1} = \frac{Q_2}{Q_1}\lp \frac{M_2}{M_1}\rp^2\lp \frac{R_1}{R_2}\rp^3 \approx 95\frac{Q_2}{Q_1},
\ee
where the term on the far right assumes the masses and radii appropriate for J0651. This result reflects the fact that the tidal torque is smaller on the C/O WD, so the asynchronicity is larger for the same $Q$. The ratio of the heating rates is
\be
	\frac{\dot{E}_2}{\dot{E}_1} = \frac{Q_2}{Q_1}\lp \frac{M_2}{M_1}\rp^3\lp \frac{R_1}{R_2}\rp \approx 29\frac{Q_2}{Q_1}.
\ee
If the observed luminosity of the C/O WD is also due to tidal heating, then $Q_2\approx (Q_1/29)(L_2/L_1)\approx 2\times10^7$. The structure of the He and C/O WDs are quite different, with the former having a shallower convective region. In future work it would be interesting to theoretically explore whether such differences lead to $Q_2\ll Q_1$.

Numerical simulations of merging WDs often assume that the WDs are not tidally locked at merger \citep[for example][]{seg97,gue04}. From our derivation we can see that this is not always the case, and in general it depends on the value of $Q$. For example, $\sigma_1/n\lesssim0.1$ merely require $Q_1\lesssim5\times10^9$. If $Q_2=2\times10^7$ right up until merger, then $\sigma_2/n\approx0.02$ and the C/O WD must be tidally locked unless $Q_2$ increases by a couple of orders of magnitude as the binary nears merger.

In exploring the survival of binary WDs in forming AM CVn systems, \citet{mar04} conclude that the synchronization time for the secondary must be less than $1,000\ {\rm yr}$. For the model presented here, that would imply $Q_2\lesssim10^8$ when accretion first begins. The current $L_2$ appears to place a more stringent limit on $Q_2$ than this already, although this conclusion again depends on how $Q_2$ changes as the orbit shrinks.

\section{Numerical Integrations}
\label{sec:calculate}

\begin{figure}
\epsscale{1.2}
\plotone{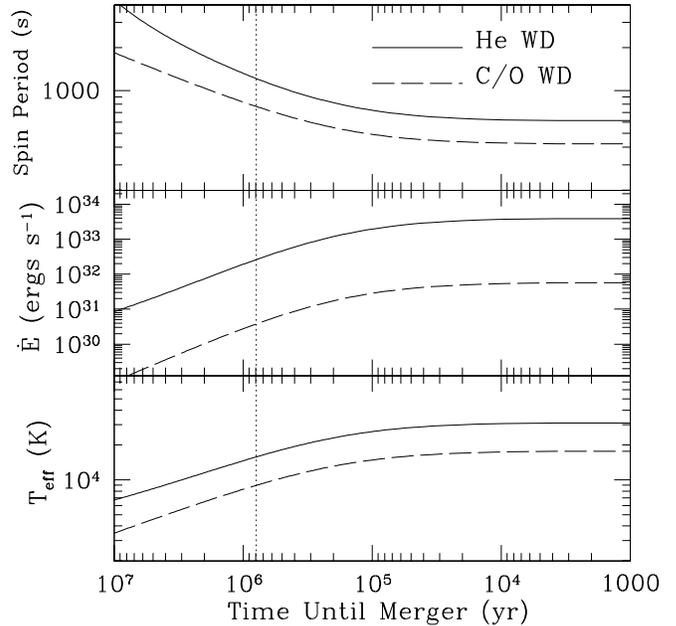}
\caption{The binary evolution as a function of time, using $Q_1=7\times10^{10}$ and $Q_2=2\times10^7$. Masses and radii are chosen to match J0651. The top panel shows the spin period of each star. The orbital period of the binary is nearly equal to the spin period of the C/O WD ({\it dashed line}) and thus is not plotted. The middle panel plots the tidal heating rate for each star, and the bottom panel shows the surface effective temperatures. The vertical dotted line shows the current location of J0651.}
\label{fig:spinevolution}
\epsscale{1.0}
\end{figure}

In Figure \ref{fig:spinevolution}, I show numerical integrations of equations~(\ref{eq:angularmomentum}) and (\ref{eq:gw}), using the mass and radii appropriate for J0651. The tidal $Q$ parameters are set to be constant at $Q_1=7\times10^{10}$ and $Q_2=2\times10^7$, so as to give heating rates at $P=765\ {\rm s}$ that are the same as the present luminosities of each star. The WDs are assumed to be non-spinning initially at a large orbital period, but are found to be quickly spun up by tides until $d\sigma/dt\approx0$ is reached, consistent with the assumptions for my analytic estimates. The integration ends when the Roche-lobe around the He WD becomes equal to its radius, which occurs at $P\approx420\ {\rm s}$ (ignoring potential changes to $R_1$ due to tidal heating). The He WD is spinning significantly more slowly than the orbital period because of its large $Q_1$. The vertical dotted line denotes the current location of J0651 at $800,000\ {\rm yr}$ before merger. If the values of $Q$ remain constant, the luminosity of the He WD increases by a factor of $\sim15$ before tidal disruption.

\begin{figure}
\epsscale{1.2}
\plotone{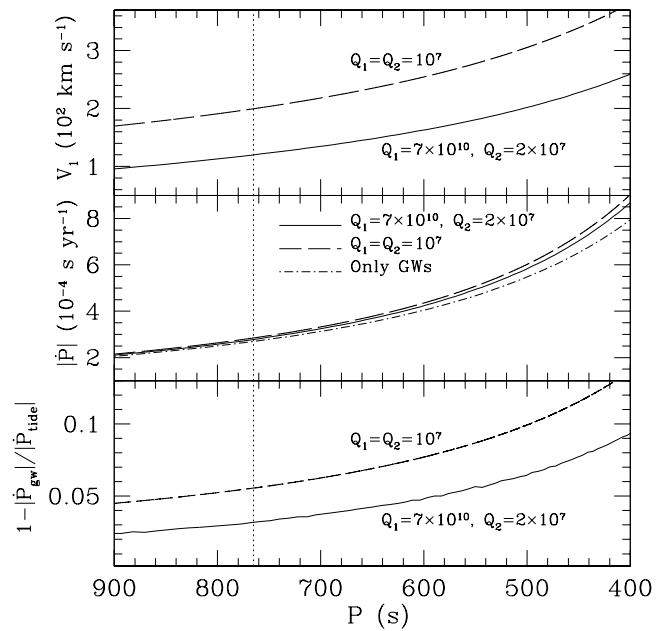}
\caption{The top panel plots the velocity of the primary $V_1=\Omega_1 R_1$ as a function of orbital period. I consider a case where the values of $Q$ are chosen to fit the observed luminosities ({\it solid line}), and a low $Q$ example where the WDs are nearly tidally locked ({\it dashed line}). The middle panel plots the period derivative, comparing merely gravitational wave losses ({\it dot-dashed line}) with the other two tidal cases. The bottom panel shows the deviation of the two tidal cases from pure gravitational wave losses. The vertical dotted line shows the current period of J0651.}
\label{fig:pdot}
\epsscale{1.0}
\end{figure}

In the top panel of Figure \ref{fig:pdot}, I calculate the rotational velocity of the primary $V_1=\Omega_1R_1$, as a function of orbital period. When the values of $Q$ are chosen to match the current luminosities, $V_1\approx 120\ {\rm km\ s^{-1}}$. Another case where $Q_1=Q_2=10^7$ is also plotted, which represents what happens when the WDs are nearly tidally locked, resulting in $V_1\approx200\ {\rm km\ s^{-1}}$. Since the binary is eclipsing, the difference between these cases may be measurable via the Rossiter-McLaughlin effect \citep{gro11}.

An important consequence of the tidal interactions is that the orbital period derivative deviates from what is expected if the system is purely driven by gravitational wave losses. Using equation (\ref{eq:taugw}), gravitational waves alone give
\be
	|\dot{P}_{\rm gw}| = 1.7\times10^{-4}M_{0.55}M_{0.25}M_{0.8}^{-1/3}P_3^{-5/3}\ {\rm s\ yr^{-1}},
	\nonumber
	\\
	\label{eq:pdotgw}
\ee
so for J0651 one would expect a period decrease of $|\dot{P}_{\rm gw}|=2.7\times10^{-4}\ {\rm s\ yr^{-1}}$. When tidal effects are included, the period decreases faster, because angular momentum is loss by the orbit to spin up the WDs. I assess this effect in the numerical models using the relation
\be
	|\dot{P}| = \frac{6\pi}{G^{2/3}}\lp \frac{P}{2\pi}\rp^{2/3}\frac{M^{1/3}}{M_1M_2}|\dot{J}_{\rm orb}|,
\ee
where $\dot{J}_{\rm orb}$ is found from equation (\ref{eq:gw}).

In the middle panel of Figure \ref{fig:pdot}, I compare the period derivative for purely gravitational wave losses ({\it dot-dashed line}) with two different degrees of tidally locking ({\it solid line} and {\it dashed line}). This confirms that the period decreases more rapidly when tidal effects are included, and this deviation increases as the binary inspirals to smaller orbital periods.

In the bottom panel of Figure \ref{fig:pdot}, I plot the fractional change in the period derivative, comparing the tidal cases with purely gravitational wave losses. For the values of $Q$ that best fit the current luminosities, the period derivative is $\approx 3.0\%$ larger, whereas if the WDs are nearly tidally locked, the period derivative changes by $\approx 5.6\%$. Looking for these deviations in future observations would provide an important measurement of the presence and level of tidal locking. An observed $|\dot{P}|\gtrsim |\dot{P}_{\rm gw}|$ implies strong tidal locking, but if $|\dot{P}|\approx|\dot{P}_{\rm gw}|$, then the asynchronicity and tidal heating must be large.

A non-zero $\dot{P}$ can also be measured by its affect on the eclipse timing. After a time $t$, this changes by
\be
	{\rm Change\ in\ eclipse\ time} = \frac{t^2}{2}\frac{ \dot{P}}{P}.
	\label{eq:eclipse}
\ee
For purely gravitational wave emission, the change in the eclipse time is $5.5\ {\rm s}$ after one year. When tidal interactions are included, this change is $3-6\%$ larger, which corresponds to eclipses  $5.6-5.8\ {\rm s}$ sooner. Associated phase shifts would be important in studies of this binary by the proposed {\it Laser Interferometer Space Antenna} mission \citep{wh98,nel04}.

\section{Discussion and Conclusion}
\label{sec:conclusion}

I considered the effect of tides in the detached binary J0651. Assuming that the current luminosity of each WD reflects its tidal heating, I found $Q_1=7\times10^{10}$ and $Q_2=2\times10^{7}$ for the He and C/O WDs, respectively. The values of $Q$ cannot be greater than this, otherwise the tidal heating would be inconsistent with the current luminosities. The degree of tidal locking can be measured in future observations, either via the Rossiter-McLaughlin effect or eclipse timing. If only gravitational waves are acting, then after one year the eclipses occur $5.5\ {\rm s}$ earlier. Tidal locking causes this difference to instead be as much as $5.8\ {\rm s}$. Since the time of eclipses scales $\propto t^2$ (see eq. [\ref{eq:eclipse}]), longer baseline observations will be important for making an accurate measurement.

Before deviations of $\dot{P}$ from merely gravitational wave losses can be inferred, the masses of the WDs must be measured sufficiently accurately. Since $\dot{P}_{\rm gw}\propto M_1M_2/M^{1/3}$ (eq.~[\ref{eq:pdotgw}]), an uncharacteristically large $|\dot{P}|$ may instead imply larger masses. Mass measurements within a couple of percent would be ideal, which in turn means that the inclination must be highly constrained. At the same time, the $\dot{P}$ cannot be strictly due to gravitational waves, otherwise the tidal heating would be too large in comparison to the observed luminosities. Once $\dot{P}$ is measured in future observations, simultaneous modeling of the masses and luminosities should be done to constrain what values of $Q$, and thus degree of tidal locking, are allowed.

A critical question about J0651 is its age and how it relates to the larger population of WD binaries. The current merger time for J0651 is $800,000\ {\rm yr}$, while the age of the He WD inferred from its current effective temperature, presuming no tidal heating, is  $\approx10^8\ {\rm yr}$ \citep{pan07}. On the face of it, this would imply that there should be $\approx100$ binaries with periods of $P(\tau_{\rm gw}=10^8\ {\rm yr})\approx50\ {\rm min}$ for every system like J0651. Although we are limited by small number statistics, such a large number of binaries has not been found \citep{kil11}. This may point to an incompleteness of the current surveys.
On the other hand, the progenitors to J0651 may be wider binaries that are older, cooler, and harder to detect. As the binary contracts, the tidal heating then makes the He WD bright and easier to observe. A more self-consistent model of the tidal heating plus stellar cooling, including changes to the He WD radius, is needed to assess what luminosity is expected as a function of time. Also, depending on when the heating takes place it can either be radiated readily or have time to heat the core and alter the stellar structure. I plan to explore these details better in a subsequent study.

Although there has been some work on estimating the $Q$ of WD binaries \citep[see][and references therein]{fl11}, more can be done, especially in the context of He WDs. The presence of a thin layer of hydrogen ($\sim10^{-3}\ M_\odot$) on the He WD will alter the eigenfunctions of the excited modes, determining which specific modes are driven as well as affecting their thermal damping timescale, which in turn determines $Q$. In addition, the lightcurves presented by \citet{bro11} for J0651 show considerable dispersion (see the upper panel of their Figure 4) that could potentially point to stellar oscillations. The hot temperature of the He WD makes it well outside of the traditional instability strip associated with WDs, which requires $T_{\rm eff}\lesssim11,000\ {\rm K}$ \citep{arr06,ste10}. But strong tidal interaction may be another method of exciting observable oscillations, which should also be investigated.

\acknowledgments
I thank Phil Arras, Lars Bildsten, Peter Goldreich, Paul Groot, Mukremin Kilic, Samaya Nissanke, Eran Ofek, Christian Ott, and Sterl Phinney for helpful feedback. This work was supported through NASA ATP grant NNX07AH06G, NSF grant AST-0855535, and by the Sherman Fairchild Foundation.


\end{document}